\documentclass[preprint,showpacs,preprintnumbers,amsmath,amssymb]{revtex4}

\usepackage{graphicx}
\usepackage{dcolumn}
\usepackage{bm}

\begin{document}

\title {SPIN-CURRENTS AND SPIN-PUMPING 
FORCES FOR SPINTRONICS}
\author{J.-E. Wegrowe} \email{jean-eric.wegrowe@polytechnique.fr}
 
\author{H.-J. Drouhin}
\affiliation{Ecole Polytechnique, LSI, CNRS and CEA/DSM/IRAMIS, 
Palaiseau 
F-91128, France.} 

\date{\today}

\begin{abstract}
A general definition of the Spintronics concept of spin-pumping 
is proposed as generalized forces conjugated 
to the spin degrees of freedom in the framework of the theory of 
mesoscopic non-equilibrium thermodynamics.  It is shown that at least 
three different kinds of spin-pumping forces and associated 
spin-currents can be defined in the most simple spintronics system; 
the Ferromagnetic/Non-Ferromagnetic metal interface.  Furthermore, the 
generalized force associated to the ferromagnetic collective variable 
is also introduced in an equal footing, in order to describe the 
coexistence of the spin of the conduction electrons (paramagnetic 
spins attached to $s$-band electrons) and the ferromagnetic-order 
parameter.  The dynamical coupling between these two kinds of magnetic degrees 
of freedom is presented, and interpreted in terms of spin-transfer 
effects.
\end{abstract}

\pacs{75.40.Gb, 72.25.Hg, 75.47.De  \hfill}

\maketitle

\section{Introduction}
Spintronics is a generalization of electronics that takes into 
account the degrees of freedom related to the spins of the conduction 
electrons (or the spin of other electric carriers). A typical spintronics system is
defined by a statistical ensemble of a plurality of electronic
populations (discriminated by their internal degrees of freedom), that
are put out of equilibrium in the presence of electric and  magnetic
forces.  The consequence is the creation of currents of electric-charge carriers and currents of spins.
Spintronics emerged with 
the discovery of giant magnetoresistance in the late 80s 
\cite{Nobel_lecture}, and it plays today a crucial role in the 
development of new electronic devices and functionalities. 

The goal of this report is revisiting spintronics on the basis of the
theory of Non-Equilibrium Thermodynamics
\cite{DeGroot,Prigogine,Stueckelberg,Smith,Parrott}.  The analysis is based
on the first and second law of thermodynamics, i.e. on the expression of
the power dissipated through the different relaxation mechanisms that
characterize the system.  The description holds at the mesoscopic
scales, under the hypothesis of local equilibrium extended to internal
degrees of freedom \cite{Prigogine0,Rubi}.  This work is restricted 
to classical systems. The extension to quantum systems, and especially 
for the definitions related to "permanent currents" (for which the 
second law of thermodynamics is inoperative) is beyond the 
scope of this report \cite{Rque}.

\subsubsection{Longitudinal spin relaxation}

The typical system to be investigated is a 1D wire (with $x$ as 
coordinate) containing a Ferromagnetic/Non-Ferromagnetic metal 
junction (or a Ferromagnetic/Ferromagnetic metal junction). In the 
most simple cases it can be described by a statistical ensemble of 
independent electrons that are defined by an effective mass $m^{*}$, 
an electric charge $e$, internal degrees of freedom reduced to the 
spin one half, and a quantization axis (fixed by the magnetization 
of the ferromagnetic layer). The system can then be reduced to two 
electronic populations locally represented by a reservoir of conduction 
electrons of spin up - defined by a chemical potential 
$\mu_{\uparrow}(x)$ - and a reservoir of conduction electrons of spin 
down - defined by a chemical potential $\mu_{\downarrow}(x)$. This is 
called the two spin-channel model.

There are then two different ways to put the system out of 
equilibrium. (i) Applying an electric field $E_{\updownarrow} = 
\frac{-1}{e} \, \frac{\partial \mu_{\updownarrow}}{\partial x}$ will create 
spin-dependent electric currents
$J^{e}_{\updownarrow} = 
\sigma_{\updownarrow} E_{\updownarrow}$  through the wire (where $\sigma_{\updownarrow}$ 
is the conductivity). (ii) applying an effective magnetic field will 
put the two spin populations out-of-equilibrium, creating in turn a 
flux of spins (noted $\dot \psi$ here) in the spin configuration 
space. 

However, the striking point is that the two generalized forces,
electric and magnetic, are not independent.  Applying a voltage
difference through the junction generates also a stationary flux of spins
$\dot \psi$ because the electric field difference $E_{\uparrow} -
E_{\downarrow} = \frac{-1}{e} \frac{\partial \Delta \mu}{\partial x}$ is
maintained in the junction.  The junction produces a non-zero chemical
potential difference $\Delta \mu = \mu_{\uparrow} - \mu_{\downarrow}$,
that plays the role of an effective magnetic field, or {\it spin-pumping
force} conjugated to the flux of spins $\dot \psi$.  The spin flux can
also be written $\dot \psi = \delta n / \tau_{\sf}$, where
$\tau_{\sf}$ is the spin-flip relaxation time and $\delta n =
n_{\uparrow} - n_{\downarrow}$ is the density of out-of-equilibrium
spins.

As a consequence, {\it applying a voltage difference through the 
junction generates not only stationary spin-dependent electric 
currents} $J^{e}_{\updownarrow} $ {\it but also the stationary spin 
flux} $\dot \psi$.

If $\mathcal{E}(\mathcal{S},n_\uparrow,n_{\downarrow})$ is the energy
density of the system (function of the entropy $\mathcal{S}$ and the
density of charge carriers $n_\uparrow$ and $n_{\downarrow}$), the
following canonical definitions holds: $ T \equiv \frac{\partial
\mathcal{E}}{\partial \mathcal{S}}$, $ \mu_{\updownarrow} \equiv
\frac{\partial \mathcal{E}}{\partial n_{\updownarrow}}$ and $\Delta
\mu \equiv \frac{\partial \mathcal{E}}{\partial \psi}$.  The first
relation defines the temperature, the second defines the chemical
potential, and the last relation {\it defines the pumping force}
$\Delta \mu$ as the chemical
affinity of the reaction that transforms a conduction electron of spin
up into a conduction electron of spin down \cite{PRB00,Revue}.

The power dissipated by the system at fixed temperature 
reads
$T d \mathcal{S}/dt = - J_{\uparrow} E_{\uparrow} - J_{\downarrow}
E_{\downarrow} + \dot \psi \Delta \mu$, where $d \mathcal{S}/dt$ is
internal entropy production.  The relation between the generalized
force and the generalized flux is imposed by the second law of
thermodynamics, $d\mathcal{S}/dt \ge 0$, and is formally expressed by
the Onsager relation.

This model has been first proposed by Johnson and Silsbee \cite{Johnson}, 
and the introduction of the pumping force $\Delta \mu$ in the context 
of spin-dependent transport is due to Van Son et al. \cite{Wyder}, 
with the description of longitudinal spin-accumulation and giant 
magnetoresistance at a 
Ferromagnetic/Non-Ferromagnetic interface. This approach was 
systematized by T. Valet and A. Fert in 1993, 
on the basis of the Boltzmann transport equations \cite{Valet}. The model is presented below 
(Section III) in the language of non-equilibrium-thermodynamics.

Note that in the Spintonics 
literature, the term Òspin-currentÓ is 
devoted to the spin-dependent electric current $\delta J^{e} = 
J^{e}_{\uparrow} - J^{e}_{\downarrow}$ that flows through the wire, 
and not to the Òspin-fluxÓ $\dot \psi$ that flows though the 
spin-space (e.g. the so-called "Bloch sphere").

\subsubsection{Spin precession}

However, the description of spin dynamics in the spin-space is not 
restricted to spin-flip relaxation (i.e. "longitudinal" spin 
relaxation) but it is also characterized by precession effects 
("transverse spin relaxation") as described e.g. by the Bloch 
equation of the paramagnetic resonance or by the Hanle effect 
measured in semiconductors \cite{Appelbaum}. The effect of the 
precession can be taken into account in terms of diffusion of the transverse 
components of the spin. This generalization was called transverse 
spin-accumulation \cite{Levy,Dugaev}, and was introduced 
much later in the context of spin-transfer experiments (see below).

As a consequence, there is another way to drive 
the spins out-of-equilibrium with the use of the transverse spin 
pumping force $\Delta \mu_{\perp}$ that generates the transverse spin 
flux $\dot \psi_{\perp}$. Also in that case, the application of a 
potential difference trough the junction gives rise to non-zero $\Delta 
\mu_{\perp} $, and the transverse 
spin-polarized current $\delta J_{\perp} = \sigma_{\perp} 
\frac{\partial \Delta \mu_{\perp}}{\partial x}$ is produced at the 
interface. The tranverse spin flux can also be expressed in terms of 
transverse relaxation time $\tau_{\perp}$ with $\dot \psi_{\perp} = 
\delta n_{\perp}/\tau_{\perp}$. These recent developments are presented in 
Section IV on equal footing with the longitudinal spin - flip.

\subsubsection{Band structure and s-d relaxation}

Furthermore, beyond the spin internal degrees of freedom, it is
important to push the description to a more realistic situation that
takes into account the ferromagnetic specificity of the material (and
not only the paramagnetic properties of the conduction electrons). 
Indeed, the out-of-equilibrium magnetization described within the
two-channel model is given by $\delta m = g \mu_B (n_{\uparrow}
- n_{\downarrow})$ (where $\mu_B$ is the Bohr magneton and $g$ the
Land\'e factor).  This system is paramagnetic.  The contribution
$\delta m$ to the total magnetization is superimposed to the
ferromagnetic collective variable $\vec M$ (which is essentially due 
to the $d$ band electrons \cite{Stearns}).  In terms of transport
properties, the quasi particles $s$ and $d$ are
defined by effective masses $m^{*}_{s}$ and $m^{*}_{d}$, i.e. by 
supplementary
internal degrees of freedom that takes into account the
coupling to the periodic lattice.

In line with the pioneering works of Mott \cite{Mott}, we will
consider a simple generalization of the two channel model that takes
into account the ferromagnetic nature of the $3d$ metals with
enlarging the internal degrees of freedom to four electronic
populations: the conduction electrons of the $s$ band for up and down
spins, and the conduction electrons of the $d$ band for up and down
spins.  We have then a four-channel model, in which two kinds of spin
flux $\dot \psi$ and $\dot \psi_{sd}$ should be defined for both
interband and intraband spin-dependent relaxation.  A third kind of
pumping force can then be introduced with the interband chemical
potential difference $\Delta \mu_{sd}$ \cite{MTEP}.  This is performed
in Section V.

\subsubsection{Ferromagnetic collective variable}

Nevertheless, there is something missing in the above description of a
ferromagnetic junction: the ferromagnetic collective variable $\vec M$
has not been introduced explicitly.  Accordingly, the next section
below (Section II) presents a derivation of the equation of the
dynamics for the ferromagnetic variable (i.e. the
Landau-Lifshitz-Gilbert equation) performed with the introduction of a
generalized force $ \vec \nabla_{\Sigma} \mu^F = - \vec{H}_{eff}$
thermodynamically conjugated to the ferromagnetic degrees of freedom
($\mu^F$ is the ferromagnetic chemical potential and $\Sigma$ is the
magnetic configuration space).  The effective magnetic field
$\vec{H}_{eff}$ can also legitimately pretend to the appellation "spin-pumping force".  This generalized ferromagnetic force generates the
current $J^{F}$ in the magnetic configuration space.

Yet, many experiments have shown that it is possible to switch the
magnetization \cite{Tsoi,EPL,Albert,Julie} or to generate ferromagnetic 
entropy \cite{Entropy} while injecting
spin-polarized currents in Ferromagnetic/Non-Ferromagnetic
junctions.  The corresponding effect - called spin-transfer - shows
that the spin-dependent electronic transport coefficients are coupled to the transport coefficient of the
ferromagnet.  In other terms, the ferromagnetic current generated by
$\vec{H}_{eff}$ and the spin-polarized current generated by $\Delta
\mu$, $\Delta \mu_{\perp}$, or $\Delta \mu_{sd}$ are coupled.

The dynamical coupling that occurs between the current of spins 
and the current of ferromagnetic moments is discussed in Sec. 
VI. It leads to define a spin-transfer effect for all 
the spin-pumping sources we have previousely identified : longitudinal 
\cite{Heide,PRB00,Revue}, transverse \cite{Sloncz,Braatas}, and s-d 
interband relaxation \cite{Berger,PRB08}. 

\section{Introduction of the ferromagnetic degrees of freedom}

In this section, we focus on a uniform ferromagnetic
moment $\vec M = M_{s} \vec u_{r}$ defined with radial unit vector
$\vec u_{r}$ and the magnetization at saturation $M_{s}$.  It is
called macrospin, in opposition to the microscopic spins attached to 
atomes or electrons.  In order to treat statistically the time
dependence of ferromagnetic degrees of freedom $\vec m$ contacted to
a heat bath, the ergodic property is used.  It allows working with a
statistical ensemble of a large number of ferromagnetic moments $\vec
m \pm d \vec m$ that defines a surface $ \{ \theta \pm \theta, \varphi
\pm d \varphi\}$ on the sphere $\Sigma$ of radius $M_{s}$.  The
corresponding density $\rho^{F}(\theta,\varphi)$ is then identified
with the statistical distribution of ferromagnetic moments
\cite{Ciornei}.  The mean value of the magnetization is $\vec M$.  
The introduction of the density is justified by the
nanoscopic size of the magnetic single domain, for which the
fluctuations play a major role.  To that point of view, the system is
mesoscopic.  Accordingly \cite{Mazur}, the ferromagnetic chemical
potential takes the general form $\mu^{F} = kT \, ln(\rho^{F}) +
V^{F}$ where $V^{F}$ is the usual ferromagnetic potential (deduced
e.g. from the quasi static hysteresis loops) and the first term
accounts for diffusion.  On the other hand the current of
ferromagnetic moments, $\vec J^{F} = \rho^F d \vec{u}_{r}/dt$, is
confined on the surface of the sphere.

 \begin{figure}
   \begin{center}
   \begin{tabular}{c}
   \includegraphics[height=7cm]{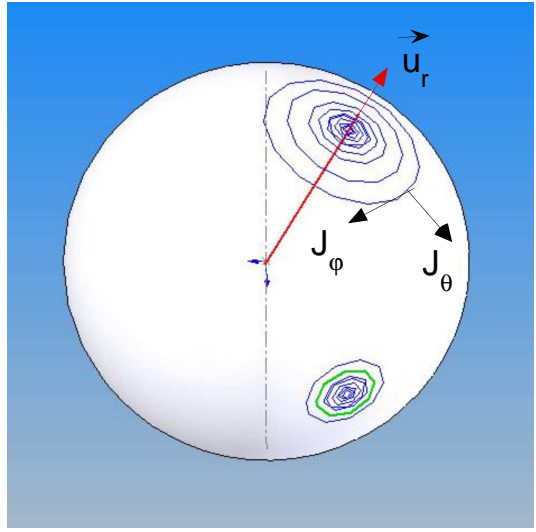}
   \end{tabular}
   \end{center}
   \caption[Ferro] 
{ \label{fig:Ferro} Illustration of the $\Sigma$ configuration space of 
the ferromagnetic variable. The two potential minima and the ferromagnetic 
currents are sketched}
   \end{figure}

The power dissipated by the ferromagnetic system is given by the 
corresponding internal entropy production $\frac{d 
\mathcal{S}^{F}}{dt}$, and is 
formed by the product of the generalized flux by the generalized 
force.  Assuming a uniform temperature $T$ we have:
 
\begin{equation}
T \frac{d\mathcal{S}^{F}}{dt} = 
- \vec J^{F} . \vec \nabla_{\Sigma} \mu^{F}
\end{equation}

The application of the second law of thermodynamics 
$d\mathcal{S}^{F}/dt \ge 0 $
allows the transport equation to be deduced
by writing the relation that links the generalized flux (the current
$\vec J^{F}$) of the extensive variables under consideration and to 
the generalized
force defined in the corresponding space $\Sigma$.  Both quantities, 
flux and
forces, are related by the Onsager matrix of the transport
coefficients $\bar{\mathcal L}$:

\begin{equation}
    \vec J^{F} = - \bar{\mathcal L} \, \vec \nabla_{\Sigma} 
    \mu^{F}
   \label{Jferro}
\end{equation}

{\it This is the simplest form of the well-known Landau-Lifshitz 
equation} (see below).
We started from the hypothesis that the magnetic domain is uniform: 
the
modulus of the magnetization is conserved.  The trajectory of the
magnetization is then confined on the
surface of a sphere of radius $M_{s}$, and the flow is a two component
vector defined with the unit vectors $\{ \vec u_{\varphi}, \vec
u_{\theta} \}$ of $\Sigma$.  Accordingly, the Onsager matrix is a 2x2 matrix
defined by four transport coefficients $\{L_{\theta \theta},
L_{\theta \varphi}, L_{\varphi \theta}, L_{\varphi \varphi} \}$. 
Furthermore, the Onsager reciprocity relations impose that $L_{\theta
\varphi} = - L_{\varphi \theta}$. Assuming that the dissipation is 
isotropic, we have $L_{\theta \theta} = L_{\varphi \varphi}$.  Let
us now introduce a dimensionless supplementary coefficient $
\alpha$,
which is the ratio of the off-diagonal to the diagonal coefficients:
$ \alpha = L_{\theta \varphi} / L_{\theta \theta} $.  In 
conclusion,
the ferromagnetic kinetic equation is defined by two ferromagnetic
transport coefficients $L_{\theta \varphi} = \rho^{F} L_{F}$ and
$ \alpha$:

\begin{equation}
\bar{\mathcal L} =
\rho^{F} L_{F} \,  \left( \begin{array}{cc}
 \alpha &  1 \\
                        - 1 & \alpha \\
\end{array} \right)
\label{MatrixL0}
\end{equation}

On the other hand, the generalized force $\vec \nabla_{\Sigma} 
\mu^{F} $, 
thermodynamically conjugated to the magnetization, is the effective magnetic field $\vec H_{eff} 
\equiv - \vec 
\nabla_{\Sigma} \mu^{F}$. It is a generalization in the sense that 
this effective 
field includes the 
diffusive term \cite{Raikher} that was first introduced by Brown 
in the 
rotational Fokker-Planck equation \cite{Brown}. 

Actually, it could be rather surprising to claim that Eq.
(\ref{Jferro}) is the "well-known
LL equation".  However, it is
sufficient to rewrite Eq. (\ref{Jferro}) in 3D space with
re-introducing the radial unit vector $ \vec u_{r} = (1,0,0) $ of the
reference frame $\{ \vec u_{r}, \vec u_{\theta}, \vec u_{\varphi} \}$,
and recalling that the current is the density multiplied by the
velocity $\vec J^{F} = \rho^{F} d\vec u_{r}/dt$, to
recover the traditional LL equation from Eq. (\ref{Jferro}) and Eq. 
(\ref{MatrixL0}):

\begin{equation}
\frac{d\vec u_{r}}{dt}  =
- L_{F} \, \left \{ 
 \vec u_{r} \times \vec H_{eff} + \alpha \vec u_{r} \times \left ( 
\vec u_{r} \times
 \vec H_{eff} \right \} 
 \right ) 
 \label{LL}
\end{equation}

Furthermore, it is well-known that LL equation is equivalent to the
phenomenological Gilbert \cite{Brown,Coffey} equation, that defines 
the 
magnetic damping
coefficient $\eta$:

\begin{equation}
\frac{d\vec M}{dt}  =
  \gamma  \vec M \times \left (
  \vec H_{eff} - \eta \, \frac{d \vec M}{dt}
 \right ) 
\end{equation}

where $\gamma$ is the gyromagnetic ratio.
The equivalence between the two equations defines the 
coefficients $\alpha$ and $L_{F}$ as a function of the coefficients 
$\eta$ 
and $\gamma$: $\alpha = \eta \gamma M_{s}$ is the dimensionless 
damping 
coefficent and $L_{F}$ is defined by the relation:

\begin{equation}
         L_{F}
		 =  \frac{\gamma}{M_{s} \left (1+ \alpha^{2} \right )}
\label{Ltheta}
\end{equation}

The above approach can be applied to 
microscopic spins (e.g. for the derivation of the Bloch equation), 
but it should be generalized to the case in which the modulus of the 
magnetization is not constant (3 X 3 matrix) and
the damping is not necessarily isotrope ($L_{\theta,\theta} \ne 
L_{\varphi,\varphi}$). This task is beyond the 
scope of this report.

\section{Two spin-channel model}

In this section, we only focus on the 
spin-dependent electric transport only, and on the two-channel model 
sketched in the introduction.  The
corresponding electric wire is defined along the $x$ axis, with a
section unity: the relevant configuration space is the one-dimensional
real space $\mathbb R$.

The conservation laws write: 

\begin{equation}
\left\{
        \begin{array}{c}
\frac{d n_{\uparrow}}{dt}\, = \,-\frac{\partial 
J^{e}_{\uparrow}}{\partial
x} - \, \dot{\psi} \\
\frac{d n_{\downarrow}}{dt}\, = \,-\frac{\partial 
J^{e}_{\downarrow}}{\partial
x}
+ \, \dot{\psi} \\
\end{array}
\label{con0}
\right.
\end{equation}

where $n_{\uparrow}$ and $n_{\downarrow}$ are the densities of charge
carriers in the channels $\{\uparrow, \downarrow \}$, and the 
spin-dependent
relaxation is taken into account by the flux $\dot \psi $.  This is
the velocity of the reaction (or relaxation of the spin-dependent
internal variable) that transforms a conduction electron $\uparrow $
into a conduction electron $\downarrow $. This relaxation is 
formally 
equivalent to a chemical
reaction, driven by the chemical affinity $\Delta \mu = 
\mu_{\uparrow} -
\mu_{\downarrow}$ \cite{PRB00}.  The power dissipated by the system then reads:

\begin{equation}
T \frac{d\mathcal{S}^{e}}{dt} =
- J_{\uparrow}^{e} \, \frac{1}{e} \frac{\partial  
\mu_{\uparrow}^{e}}{\partial 
x} -  J_{\downarrow}^{e} \,  \frac{1}{e} \frac{\partial  
\mu_{\downarrow}^{e}}{\partial 
x} + \dot \psi \Delta \mu
\label{PowerGMR}
\end{equation}

The corresponding kinetic equations are deduced from the second law of
thermodynamics, after introducing a supplementary Onsager coefficient
$L$.

\begin{equation}
\left\{
\begin{aligned}
J^{e}_{\uparrow } &= -\frac{\sigma_{\uparrow}}{e} \frac{\partial
\mu^{e}_{\uparrow}}{\partial x} \\
J^{e}_{\downarrow } &= -\frac{\sigma_{\downarrow}}{e} \frac{\partial
\mu^{e}_{\downarrow}}{\partial x} \\
\dot \psi &= L \Delta \mu
\end{aligned}
\right. \label{OnsagerGMR}
\end{equation}

The set of equations Eqs (\ref{OnsagerGMR}) is sufficient and
necessary in order to describe, in the stationary regime, 
spin-accumulation 
effects and any
non-equilibrium contribution to the resistance due to relaxation 
($\Delta \mu \ne 0$) occurring at an
interface \cite{Revue,MTEP}. 

Eq. (\ref{OnsagerGMR}) shows that the spin-dependent electric 
currents $J^{e}_{\updownarrow}$ and the spin flux $\dot \psi$ are not 
independent. 
Accordingly, it is more convenient
to rewrite Eq. (\ref{OnsagerGMR}) as a function of the variables 
$\Delta \mu$ and $\mu_{0} = \mu_{\uparrow} + \mu_{\downarrow}$. 
Let us 
define  the conductivity asymmetry by the parameter $\beta$ such
that $\beta = \frac{\sigma_{\uparrow} - 
\sigma_{\downarrow}}{\sigma_{0}}$ 
and
the mean conductivity $ 2 \sigma_{0}= \sigma_{\uparrow} + 
\sigma_{\downarrow}$.  On the other hand, the
spin-polarized electric current is $\delta J^{e} = J^{e}_{\uparrow} -
J^{e}_{\downarrow} $ and the spin-independent current is $J^{e}_{0} = 
J^{e}_{\uparrow} + J^{e}_{\downarrow} $.  In this new system of 
equations, 
the Onsager matrix
re-writes:

\begin{equation}
\left( \begin{array}{c}
\delta J^e \\
J^{e}_{0} \\
\dot \psi 
\end{array} \right)
=  \left( \begin{array}{ccc}
                        \sigma_0 & \beta \sigma_0 & 0\\
                       \beta \sigma_0 & \sigma_0 & 0 \\
			 0 & 0 & L
                        \end{array} \right)
                        \left( \begin{array}{c}
\frac{-1}{e} \frac {\partial{\Delta 
{\mu}^{e}}}{\partial {x}}\\
\frac{-1}{e} \frac{\partial \mu^{e}_{0}}{\partial x}\\
\Delta \mu \\
 \end{array} \right)
\label{OnsagerElec00}
\end{equation}

The current $\delta J^{e}$ is called "spin current" 
or "pure spin-current" in the spintronics literature.
 
The system of equations Eq.  (\ref{OnsagerElec00}) allows the 
diffusion
equation for $\Delta \mu(x)$ to be derived for the stationary 
conditions $ \frac{\partial J_{0}^{e}}{\partial x} = 0$ and 
$\frac{\partial 
\delta J^{e}}{\partial x }= - 2 \dot \psi$:

\begin{equation}
\frac{ \partial^{2} \Delta \mu}{\partial x^{2}} = \frac{\Delta 
\mu}{l^{2}_{diff}} 
\label{Diff}
 \end{equation}
 
 where $l_{diff}^{-2} = \frac{2eL}{\sigma_{0}(1-\beta^{2})}$.  And the
 non-equilibrium magnetoresistance produced by the interface writes
 (see the details of the derivation in references
 \cite{Revue,MTEP,PRB08}):
 
\begin{equation}
      R^{ne}= -\frac{2 \beta}{J_{0}^{e}e} \int_{A}^{B}
\frac{\partial \Delta \mu}{ \partial x}dx 
\label{Rneq}
\end{equation}

where the measurement points $A$ and $B$ are located far enough in
each side of the interface so that $\Delta \mu^{e}
(A)=\Delta \mu^{e} (B) =0$.

 \begin{figure}
   \begin{center}
   \begin{tabular}{c}
   \includegraphics[height=7cm]{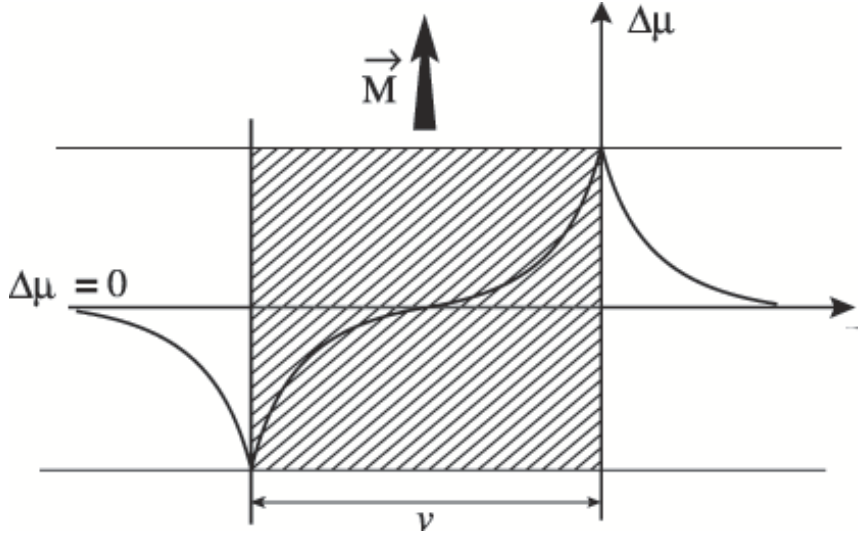}
   \end{tabular}
   \end{center}
   \caption[Spin] 
{ \label{fig:Spin} Spin-accumulation $\Delta \mu (x)$ trough
a Non-Ferromagnetic/Ferromagnetic/Non-Ferromagntic junction with 
typical size of the order of the spin diffusion length}
   \end{figure}

The spin diffusion length is typically some tens of nanometers in 
ferromagnetic metals, so that the non-equilibrium magnetoresistance 
requires thin films (current-in-plane geometry) or nanostructured 
pillars (current-perpendicular-to-the-plane geometry) to be exploited.

\section{Spin precession}

The description proposed above with a spin-dependent internal 
variable that takes
the two spin values $\{\uparrow, \downarrow \}$ is not able to take
into account the precession of the spins occurring in a magnetic 
field, and observed with electronic resonance or Hanle effects. 
In the case of the processes
that lead to spin accumulation and giant magnetoresistance, the mean 
values are
averaged out over the spin-diffusion length, so that the precession 
of the spin is not relevant. However, this is no longer the case
in a quasi-ballistic regime close enough to the interface.

In order to take into account these quasi-ballistic effects (i.e. 
sub-nanometric scales in metalic devices),
the two-channel model has been recently generalized to transverse
spin-accumulation in the context of spin-transfer-torque
investigations \cite{LevyFert,Levy,Dugaev}. 
The transverse spin-accumulation is introduced with the corresponding
current $\delta J^e_{\perp}$ and the corresponding chemical potential
$\Delta {\mu}^{e}_{\perp}$. Transverse means here that the spin 
density is considered in the plan perpendicular to the 
quantification axis $\updownarrow$ that defines the spin up and spin 
down in the two-channel model.

The conservation laws writes:

\begin{equation}
    \frac{\partial \delta n_{\perp}}{\partial t} = - 
\frac{\partial J_{\perp}}{\partial x} - \dot \psi_{\perp}
\label{ConsTransverse}
\end{equation}

The transverse spin flux $\dot \psi_{\perp}$ can be expressed with a
transverse relaxation time $\tau_{\perp}$: $\dot \psi_{\perp} =
\frac{\delta n_{\perp}}{\tau_{\perp}} $, where $n_{\perp}$ is the
density of transverse spins.

Note
that the two potentials $\Delta \mu^{e}$ and $\Delta \mu_{\perp}^{e}$
are defined at very different length scales and it is necessary to
refer to quantum approaches in order to understand the physical
signification of the transverse parameters \cite{Waintal,Braatas}. 
The corresponding transverse contribution to the dissipated power is

\begin{equation}
T \frac{d\mathcal{S}^{e}_{\perp}}{dt} = - \delta J^{e}_{\perp} \, 
\frac{\partial 
\Delta \mu^{e}_{\perp}}{e \partial
x} + \dot \psi_{\perp} \Delta \mu_{\perp}
\label{entropy}
\end{equation}

Putting all together, we have the following Onsager relations for the 
electric 
system:

\begin{equation}
\left( \begin{array}{c}
\delta J^e \\
J^{e}_{0} \\
\delta J^{e}_{\perp}\\
\dot \psi \\
\dot \psi_{\perp}
\end{array} \right)
=  \left( \begin{array}{ccccc}
                       \sigma_0 & \beta \sigma_0 & 0 & 0 & 0\\
                       \beta \sigma_0 & \sigma_0 & 0 & 0 & 0\\
		        0  &   0  &  \sigma_{\perp} &  0 & 0 \\
			 0 & 0 & 0 & L & 0 \\
			0 & 0 & 0 & 0 & L_{\perp}
                        \end{array} \right)
                        \left( \begin{array}{c}			
\frac{-1}{e} \frac {\partial{\Delta 
{\mu}^{e}}}{\partial {x}}\\
\frac{-1}{e} \frac{\partial \mu^{e}_{0}}{\partial x}\\
\frac{-1}{e} \frac {\partial{\Delta 
{\mu}^{e}_{\perp}}}{\partial {x}}\\
\Delta \mu \\
\Delta \mu_{\perp} \\
 \end{array} \right)
\label{OnsagerElec2}
\end{equation}

\section{The role of the $d \downarrow$ electronic subband}

A justification of the spin-dependent conductivity asymmetry $\beta 
\ne 0$ in 
ferromagnetic metals has been proposed by N. Mott in 1936 \cite{Mott}, 
on the bases of the newly discovered band-structure approach. In the 
Mott description, the observed transport properties (e.g. the huge 
resistivity of Ni below the Curie temperature) have been 
accounted for by the existence of four electronic populations: the 
conduction electrons of spin up and down ($ s \uparrow$ and $ s
\downarrow$) of the $s$ band and the 
conduction electrons of spin up and down ($d \uparrow$ and $d
\downarrow$) of the $d$ band. The 
argument is based on the fact that the contribution to the 
resistivity of the $s-d$ 
interband scattering is higher than the contribution of the $s$ 
intraband scattering. In the 
ferromagnetic 3d metal, the $d \uparrow$ band is full so that the 
relaxation channel of $s \uparrow$
electrons to $d \uparrow$ band is blocked (according to 
the Fermi golden rule, the relaxation rate is proportional to the 
density of states in the final $d \downarrow$ band). Furthermore, 
the 
spin-flip interband relaxation is too energetic to be efficient (the 
relaxation $s \uparrow$ to $d \downarrow$ is negligible). As a 
consequence, the 
$s \downarrow$ are more scattered that the $s \uparrow$, and 
the conductivities of the two channels is asymmetric: 
$\sigma_{\uparrow} > \sigma_{\downarrow}$. This mechanism is also 
responsible for the anisotropic magnetoresistance \cite{Potter}.
The necessity of enlarging the internal degrees of freedom to the 
band structure leads to enrich the concept of spin-currents and the 
spin-pumping force.

Using the notations introduced in the previous sections, the total
current $J_{t}$ is composed by the three currents for each channel :
$J_{t} = J_{s \uparrow}+J_{s \downarrow}+J_{d \downarrow}$ ($J_{d
\uparrow} = 0$ because the band is full).  The relaxation rate $\dot
\psi_{sd} $ is introduced to account for $s-d$ spin-conserved
scattering, and the relaxation rate $\dot \psi_{s} $, is introduced in
order to account for previously defined spin-flip scattering. 
Assuming that all channels are in steady states, the 
conservation law write:

\begin{equation}
\left\{
\begin{aligned}
          \frac{\partial n_{t}}{\partial t}& =  - \frac{\partial
J_{t}}{\partial x} = 0 \\
\frac{\partial n_{s \uparrow}}{\partial t}\, &= \,-\frac{\partial J_{s
\uparrow}}{\partial
x} - \, \dot{\psi}_{s} = 0 \\
\frac{\partial n_{s \downarrow}}{\partial t}\,& = \,-\frac{\partial 
J_{s
\downarrow}}{\partial x} - \, \dot{\psi}_{sd}
         + \, \dot{\psi}_{s} = 0\\
         \frac{\partial n_{d \downarrow}}{\partial t}\,&  =
-\frac{\partial J_{d
\downarrow}}{\partial x} + \dot{\psi}_{sd} = 0
\end{aligned}
\right. \label{con}
\end{equation}

where $n_{t}, n_{s \uparrow}, n_{s \downarrow}, n_{d \downarrow}$
are respectively the total densities of particles and the density of
particles in the $s \uparrow$, $s \downarrow$, $d
\downarrow$ channels. The conjugate
(intensive) variables are the chemical potentials $\{
\mu_{s\uparrow}, \mu_{s \downarrow}, \mu_{d \uparrow}, \mu_{d
\downarrow }\}$. The application of the
first and second laws of thermodynamics allows us to
deduce the
Onsager relations of the system :

\begin{equation}
\left\{
\begin{aligned}
J_{s \downarrow} &= -\frac{\sigma_{s \downarrow}}{e} \frac{\partial
\mu_{s \downarrow}}{\partial x}\\
J_{s \uparrow} &= -\frac{\sigma_{s \uparrow}}{e} \frac{\partial
\mu_{s
\uparrow}}{\partial x}\\
J_{d \downarrow}& = -\frac{\sigma_{d \downarrow}}{e} \frac{\partial
\mu_{d \downarrow}}{\partial x}\\
\dot{\psi}_{sd} &= L_{sd} \left ( \mu_{s
\downarrow}-\mu_{d \downarrow} \right ) \\
\dot{\psi}_{s}& = L_{s} \left ( \mu_{s \uparrow}-\mu_{s  \downarrow}
\right )
\end{aligned}
\right. \label{OnsagerF0}
\end{equation}

where the conductivity of each channel $ \{ \sigma_{s \uparrow},
\sigma_{s \downarrow}, \sigma_{d \uparrow}, \sigma_{d \downarrow}
\}$ has been introduced. The first four equations are the
Ohm's law applied to each channel, and the two last equations
introduce new Onsager transport coefficients,
$L_{sd \downarrow}$ and $L_{s}$, that respectively describe the
$s-d$ relaxation (I) for minority spins under the action of the
chemical potential difference $\Delta \mu_{\downarrow} = \mu_{s
\downarrow}/2-\mu_{d \downarrow}$ and the spin-flip relaxation (II)
under spin pumping $\Delta \mu_{s} = \mu_{s \uparrow}-\mu_{s
\downarrow}/2$. The Onsager coefficients are
proportional to the corresponding relaxation times \cite{Revue}.

In the same manner as performed in Section III, the equations of
conservation Eqs.~(\ref{con}) and the Onsager equations
Eqs.~(\ref{OnsagerF0}) lead to the two coupled diffusion equations :

\begin{equation}
\left\{
\begin{aligned}
\frac{\partial^2 \Delta \mu_{\downarrow}}{\partial x^2}\,&=\,
\frac{1}{l_{sd}^2} \, \Delta
\mu_{\downarrow}- \frac{1}{\lambda_{s}^{2}} \Delta \mu_{s} \\
\frac{\partial^2 \Delta \mu_{s}}{\partial x^2}\,&=\,
\frac{1}{\lambda_{sd}^2} \, \Delta \mu_{\downarrow}-
\frac{1}{l_{sf}^2} \, \Delta \mu_{s}
\end{aligned}
       \right.
       \label{diff}
\end{equation}

where the four diffusion lengths $\{ l_{sd}, \lambda_{s},l_{sf},
\lambda_{sd}\}$ are given as a function of the transport coefficients
in reference \cite{Revue}.

This three-channel model brings to light the interplay between band
mismatch effects and spin accumulation, in a diffusive approach. 
The resolution of the coupled diffusion equations is discussed
elsewhere \cite{Revue}.


\section{Derivation of Spin transfer due to spin-pumping forces}

In usual experimental configurations for spin-transfer, an electric
current is injected in a ferromagnet through an interface (in
series or in non-local configuration
\cite{Otani}) and the magnetoresistance, i.e. the potential drop Eq.  
(\ref{Rneq}) allows the
magnetization states to be measured. The effect of strong electric 
currents on the magnetization
states can then be observed.  In such a configuration, the two 
sub-systems $\Sigma$ and the spin-polarized current
described in the previous sections exchange magnetic moments at the
junctions and both are open systems.

In order to describe the dynamics of the ferromagnetic degrees of
freedom, we have to
deal with a closed system.  The system of interest is now the
ferromagnetic system that includes spin-accumulation effects at the
junctions.

For the sake of simplicity, we treat in the following a unique decoupled 
spin-dependent process 
$\Delta \mu$ and $\dot \psi$ that includes $s-d$ relaxation.

This total ferromagnetic system is such that the density of 
ferromagnetic moments $\rho^{F}_{tot}$ and the total ferromagnetic 
flux $\vec 
J^{F}_{tot}$ are related by the conservation law:  $d 
\rho^{F}_{tot}/dt = -
div_{\Sigma} \vec J^{F}_{tot}$.

The initial
configuration space of magnetic moments is then extended
to 1D real space parametrized by the internal
variable $\Sigma \otimes \mathbb R_{\Sigma}$. The
important point here is that the internal variable is spin dependent, 
and
related to the ferromagnetic space $\Sigma$ (e.g. through spin-flip 
or $s-d$ 
relaxation and the corresponding spin accumulation). This accounts 
for the 
coupling, i.e. the {\it transfer}, of magnetic moments between the two 
sub-systems.

The dissipation is given by the internal power dissipated in the 
total system 
$ T \, d \mathcal{S}/dt$ :

\begin{equation}
T \, d \mathcal{S}/dt = - \vec{j}_{tot}^F . \vec 
{\nabla}_{\Sigma} \mu
^F - \delta J^e . \frac {\partial{\Delta {\mu}^e}} {e \partial 
{x}} - \delta J^e_{\perp} . \frac {\partial{\Delta {\mu}^e_{\perp}}} 
{e \partial 
{x}} -
J_{0}^e \frac { \partial{\mu_{0}^e}} {e \partial {x}} + \dot \psi 
\Delta \mu^{e} + \dot \psi \Delta \mu^{e}_{\perp}
\label{source}
\end{equation}

Where the first term in the right hand side is the power dissipated 
by 
the total ferromagnetic sub-system (including the ferromagnetic 
contribution due 
to spin-transfer), the two following terms are the 
power dissipated by spin-dependent electric transport, and the fourth 
term 
is the spin-independent Joule 
heating. The last term is the power dissipated by spin-flip
or $s-d$ relaxation.

In Eq.  (\ref{source}), the vectors are defined on the
sphere $\Sigma$ with the help of two angles $\theta$ and $\varphi$. 
The total ferromagnetic current $\vec{j}_{tot}^F = j_{tot \, \perp}^{F}
\vec u_{\varphi} + j_{tot \, //}^{F} \vec u_{\theta} $ includes the
contribution due to spin-accumulation mechanisms. The chemical 
potential
$\mu^{F}$ accounts for the energy of a ferromagnetic layer. On the
other hand, the system is contacted to electric reservoirs with the 
electric currents and the corresponding chemical potentials.
Applying the second law of thermodynamics, we obtain the general 
Onsager relations:

\begin{equation}
\left( \begin{array}{c}
j_{tot \, \perp}^{F} \\
j_{tot \, //}^{F} \\
\delta J^e_{\perp} \\
\delta J^e \\
\dot \psi_{\perp}  \\
\dot \psi
\end{array} \right)
=   \left( \begin{array}{cccccc}
                       \rho L_{F} \alpha & \rho L_{F} & 
                      l_{\perp} & 0 & 0 & 0\\
		       - \rho L_{F} & \rho L_{F}  \alpha &  0 & l_{//} & 0 & 0\\
			\tilde l_{\perp} & 0 & 
\sigma_{\perp} & 0  & 0 & 0 \\
                        0 & \tilde l_{//} & 0 & \sigma_0 & 0  & 0\\
                        0 & 0 & 0 & 0 & \L_{\perp} &  0 \\
 0 & 0 & 0 & 0 & 0 & L 
\end{array} 
\right) 
\left( \begin{array}{c}
 \frac{1}{sin \theta} \frac{\partial \mu_{\perp}^F}{\partial \varphi} \\
  \frac{\partial \mu_{//}^F}{\partial \theta} \\
 \frac{-1}{e} \frac {\partial{\Delta {\mu}^{e}_{\perp}}}{\partial {x}} \\
\frac{-1}{e} \frac {\partial{\Delta {\mu}^{e}}}{\partial {x}}\\
\Delta \mu_{\perp}\\
\Delta \mu \\
 \end{array} \right)
\label{Onsagermatrix}
\end{equation}

All coefficients were defined in the previous sections, except the two cross-coefficients $ \{ l_{\perp}, l_{//} \}$, introduced in this model as spin-transfer coefficients.  The coefficients $\{\tilde l_{\perp}, \tilde 
l_{//} \}$ are given by the Onsager reciprocity relations. We assumed that the other cross-coefficents are zero or negligible.

The total ferromagnetic current can be written after integrating over
the volume $v$ of the ferromagnetic layer of section unity and the
spin accumulation zone.  This volume is such that $v = \int_{A}^{B}
dz$, where $x = A$ and $x = B$ are two sections close to the interface
but far enough with respect to the diffusion lengths.  We assume here
that the diffusion lengths are much smaller than the width of the
ferromagnetic layer in order to simplify the calculation: the volume
of the ferromagnet is identified as $v$.  Let us define $ \vec X$ as
the correction due to the spin-transfer after
integrating over the volume $v$ deduced from the two first
equations of the matrix equation Eq.  (\ref{Onsagermatrix}),:

\begin{equation}
\vec J_{tot}^{F} =  \bar{\mathcal L} v \, \vec \nabla_{\Sigma} \mu^{F} + \rho \, v \vec X 
\end{equation}

where $\bar{\mathcal L}$ is the matrix defined in Eq. 
(\ref{MatrixL0}).

The assumption of constant modulus of the magnetization imposes 
that $\vec X $ is confined on the surface of the sphere $\Sigma$. The 
Helmoltz decomposition theorem can then be applied: the vector $\vec X$  can be 
decomposed in a unique way with the introduction of two potentials 
$\chi$ and $\Phi$ 
(i.e. a 
potential vector) such that:

\begin{equation}
\vec X = \vec u_{r} \times \vec \nabla_{\Sigma} \Phi + \vec 
\nabla_{\Sigma} \chi
\label{Helmoltz}
\end{equation}

where the first term is divergenceless and the second term is 
curlless (i. e. non conservative). The total correction to the Landau-Lifshitz-Gilbert equation
    writes

\begin{equation}
\vec J_{tot}^{F} = v \bar{\mathcal L} \vec \nabla_{\Sigma} \mu^{F} + \vec 
\nabla_{\Sigma} \chi + \vec u_{r} \times \vec \nabla_{\Sigma} \Phi
\label{mutot}
\end{equation}

The generalized LLG takes the form:
\begin{equation}
\frac{d\vec{u}_{r}}{dt}  =    - L_{F}
      \left \{ 
 \vec{u}_{r}\times \left (\vec{H}_{eff} - \frac{\vec \nabla_{\Sigma} \Phi}{L_F}  \right) 
 -  \alpha \vec{u}_{r} \times \left( 
			   \vec{u}_{r} \times  \left ( \vec{H}_{eff} - \frac{ \vec \nabla_{\Sigma}{\chi}}{\alpha L_F} \right ) \right) 
\right \}
\label{GenLLG3}
\end{equation}

where $ \vec H_{eff} = - \vec \nabla \mu^{F}$ is the usual effective ferromagnetic field. 

Eq. (\ref{GenLLG3}) is a generalized LLG equation that includes the effect of spin-pumping $\Delta \mu$ and $\Delta \mu_{\perp}$ (the contribution of $\Delta \mu_{sd}$ has not been added to the Onsager relations Eq. (\ref{Onsagermatrix}) for the sake of simplicity). Qualitiatively, the most important point is that the dynamics should now be described by the introduction of two potentials, or two magnetic fields $\vec{H}_{eff} - \frac{\vec \nabla_{\Sigma} \Phi}{L_F}$ for the precession and $ \vec{H}_{eff} - \frac{ \vec \nabla_{\Sigma}{\chi}}{\alpha L_F}$ for the longitudinal relaxation, instead of a single field $\vec{H}_{eff}$ in the usual case. Both fields appearing in the LLG equation contains a current dependence.

The corresponding Fokker-Planck equation is obtained by inserting the
expression of $\vec J_{tot}$ into the conservation equation: $ d
\rho^{F}_{tot}/dt = - div_{\Sigma} \vec J^{F}_{tot}$.

\section{Conclusion}

The basic concepts of Spintronics - spin-pumping force and spin 
currents -
have been presented on the basis of the theory of Non-Equilibrium
Thermodynamics.  It has been shown that different relaxation
mechanisms can be invoked, each of which defines a specific spin-pumping force 
and a specific spin flux. Four mechanisms have been 
investigated within this formalism: spin-flip scattering, spin precession, $s-d$ 
scattering, and ferromagnetic relaxation (with both longitudinal relaxation and 
precessional motion). 

This approach shows that the application of a voltage difference 
through a Ferromagnetic/Non-Ferromagnetic junction leads to the 
creation of spin-pumping and spin-flux, which in turn leads to 
non-equilibrium interface resistance or spin-transfer to the ferromagnetic 
collective variable. 
The derivation proposed for the spin-accumulation formulae, dynamics of the magnetization, and spin-transfer effects, are based on the expression of the entropy production and the second law of thermodynamics. In principle, it is possible to obtain the same results in a more elegant way, by using the PrigogineÕs theorem on minimal entropy production. This work rest to be performed.


\bibliographystyle{mdpi}
\makeatletter
\renewcommand\@biblabel[1]{#1. }
\makeatother

\end{document}